\documentclass[aps]{revtex4}
\def\pochh #1#2{{(#1)\raise-4pt\hbox{$\scriptstyle#2$}}}
\def\binom#1#2{\left(\begin{array}{c}#1\\#2\end{array}\right)}
\font\mas=msbm10 \font\mass=msbm7 \textfont8=\mas \scriptfont8=\mass
\def\masy{\fam8}
\def\ms#1{{\masy#1}}
\usepackage{graphicx} 
\usepackage{dcolumn} 

\begin {document}

\title {Functional integral with $\varphi^4$ term in the action \\ beyond
standard perturbative methods II}
\author{J. Boh\' a\v cik}
\email{bohacik@savba.sk} \affiliation{Institute of Physics, Slovak
Academy of Sciences, D\' ubravsk\' a cesta 9, 845 11 Bratislava,
Slovakia.}
\author{P. Pre\v snajder}
\email{presnajder@fmph.uniba.sk} \affiliation{Department of
Theoretical Physics and Physics Education, Faculty of Mathematics,
Physics and Informatics, Comenius University, Mlynsk\' a dolina F2,
842 48 Bratislava, Slovakia.}

\begin{abstract}
To avoid problems with infinite measure, the functional integral for
harmonic oscillator can be calculated by time - slicing method with
continuum limit procedure proposed Gelfand and Yaglom. In previous
article we proved by nonperturbative calculation the generalized
Gelfand-Yaglom equation for anharmonic oscillator with positive or
negative mass term. In this article we prove by step-by-step the
calculation of the correction function to the Gelfand-Yaglom
equation for an-harmonic oscillator.
\end{abstract}
\maketitle

\section*{Introduction}

Let us demonstrate the meaning of continuum limit procedure proposed
by Gel'fand-Yaglom  on the example of the harmonic oscillator. In
Euclidean variant of the theory the continuum functional integral
for harmonic oscillator can be read:

$$\mathcal{Z} = \int [\mathcal{D}\varphi(x)]\exp (-\mathcal{S})\ $$

Where the euclidean action is:

\begin {equation}
\mathcal{S} =\int \limits _0^\beta d\tau \left[c/2
\left(\frac{\partial\varphi(\tau)}{\partial\tau}\right)^2+b\varphi(\tau)^2\right]
 \nonumber
\end {equation}

In time-slicing approximation of Wiener unconditional measure
functional integral we calculate the $N-$ dimensional integral
\cite{dem}:

\begin {equation}
\mathcal{Z}^0_{N}=\int\limits _{-\infty}^{+\infty} \prod \limits
_{i=1}^N\left(\frac{d\varphi_i}{\sqrt{\frac{2\pi\triangle}{c}}}\right)
\exp\left\{-\sum\limits _{i=1}^N \triangle\left[c/2
\left(\frac{\varphi_i-\varphi_{i-1}}{\triangle}\right)^2
+b\varphi_i^2\right]\right\} \nonumber
\label{int1}
\end {equation}

 where $\triangle=\beta/N$, and $b, c$ are the parameters of the model. The unconditional measure
 $N$ dimensional approximation (fixed $\varphi_0$
 and integration over $\varphi_N$ ) can be evaluated explicitly:

\begin {equation}
\mathcal{Z}^0_{N} = \left[\prod_{i=0}^{N}2(1+b\triangle^2/c)
\omega_i\right]^{-\frac{1}{2}}  \nonumber
\end {equation}

\noindent where $\omega_i$ is defined by recursion $$\omega_i =
1-\frac{A^2}{\omega_{i-1}}$$ with the first term
$$\omega_0 \; =\; 1/2 + \frac{b\triangle^2/c}{2(1+b\triangle^2/c)},$$ where
 $$A=\frac{1}{2(1+b\triangle^2/c)}$$

The recurrence relation  for functional integral is introduced by
time -- slicing procedure. Explicitly this recurrence is represented
by factor $\omega_i$. Following this recurrence, we can prove the
difference equation for inverse square root of the finite
dimensional integral. The continuum limit of this difference
equation give the non-trivial value for functional integral. Based
on this procedure the continuum limit $N\rightarrow \infty$ of the
$\mathcal{Z}^0_{N}$ can be defined by:

$$\mathcal{Z}^0(\beta)=\frac{1}{\sqrt{F(\beta)}}.$$

It was shown by  Gelfand and Yaglom that  $F(\beta)$ is the solution
of the equation:

\begin {equation}
\frac{\partial^2}{\partial \tau^2}F(\tau) = \frac{2b}{c}\; F(\tau)\;
,\; \tau \in(0, \beta). \nonumber\label{int2}
\end {equation}

For the harmonic oscillator the equation (\ref{int2}) can be
calculated by taking the continuum limit of the difference equation
extracted from the recurrence relation for $\mathcal{Z}_{N}^0.$

 The general solution of the above equation is:
\begin {equation}
F(\tau) = C_1 \cosh{\left(\sqrt{\frac{2b}{c}}\; \tau
\right)}+C_2\sinh{\left(\sqrt{\frac{2b}{c}}\; \tau \right)}
\nonumber
\end {equation}

\noindent where $C_1, C_2$ are the constants fixed by initial
conditions of the unconditional (when $\varphi_0$ is fixed and
$\varphi_N$ is free), or conditional (when both $\varphi_0,
\varphi_N$ are fixed) measure functional integral. From the
analytical form of the result for functional integral follows the
clear interpretations of the calculated quantities as energy levels
and others.

In this article we report on the attempt to evaluate
non-perturbatively the functional integral for an-harmonic
oscillator with positive as well as negative mass squared term. Our
aim is  to calculate the $N-$ dimensional integral by another method
as well-known conventional perturbative calculation. We find the
result suitable  to evaluate by the recurrence procedure the
difference equation. Following the idea of Gelfand - Yaglom we
define the differential, Gelfand - Yaglom type equation for the
quantity $y(\tau)$ related (similarly as $F(\tau)$) to the
unconditional measure functional integral. The differential
equations for $y(\tau)$ reads:

\begin {equation}
\frac{\partial^2}{\partial \tau^2}y(\tau)+4\frac{\partial}{\partial
\tau}y(\tau)\ \frac{\partial}{\partial \tau}\ln{S(\tau)} =
y(\tau)\left(\frac{2b}{c}-2\frac{\partial^2}{\partial
\tau^2}\ln{S(\tau)}-4(\frac{\partial}{\partial
\tau}\ln{S(\tau)})^2\right) , \nonumber
\end {equation}

We shall evaluate the function $S(\tau)$ analytically  in this
article.

As we explain in the text, our result possesses the form of an
asymptotic expansion power series. Nevertheless we evaluate
precisely all finite difference mathematical objects. In the article
\cite{paper1} quoted as "article I" and in this article we explain
and prove all analytical calculations.

For an-harmonic oscillator the correction term is
 $-2\frac{\partial^2}{\partial
\tau^2}\ln{S(\tau)}-4(\frac{\partial}{\partial \tau}\ln{S(\tau)})^2$
to the equation for the harmonic oscillator. It would be desirable
to recover it using more general arguments and to see its form for
some more general classes of potentials.

This article is organized as follows. In the next section we resume
the article I. In the third section we evaluate the function
$S(\tau)$ by recurrence relation from the result of $N-$ dimensional
integral. In the fourth section we discuss some preliminary
conclusions of our calculation.

\vspace{1.5cm}

\section*{Resume of the article I.}

Our aim is to solve the problem of evaluation of the continuum
unconditional measure Wiener functional integral:
$$\mathcal{Z} = \int [\mathcal{D}\varphi(x)]\exp (-\mathcal{S})\ ,$$
where continuum action possesses the fourth order term:
\begin {equation}
\mathcal{S} =\int \limits _0^\beta d\tau \left[c/2
\left(\frac{\partial\varphi(\tau)}{\partial\tau}\right)^2+b\varphi(\tau)^2
+a\varphi(\tau)^4\right]\ . \nonumber
\end {equation}

The functional integral $\mathcal{Z}$ is defined by limiting
procedure of the finite dimensional integral $\mathcal{Z}_{N}$:

\begin {equation}
\mathcal{Z}_{N}=\int\limits _{-\infty}^{+\infty} \prod \limits
_{i=1}^N\left(\frac{d\varphi_i}{\sqrt{\frac{2\pi\triangle}{c}}}\right)
\exp\left\{-\sum\limits _{i=1}^N \triangle\left[c/2
\left(\frac{\varphi_i-\varphi_{i-1}}{\triangle}\right)^2
+b\varphi_i^2+a\varphi_i^4\right]\right\} ,
\label{findim}
\end {equation}
\noindent
 where $\triangle=\beta/N$. Then, the continuum Wiener unconditional measure functional integral is
defined by the formal limit:
$$\mathcal{Z} = \lim_{N\rightarrow \infty}\;\mathcal{Z}_{N}\ .$$

The first important task is to calculate the one dimensional
integral
\begin {equation}
I_1=\int\limits _{-\infty}^{+\infty}\;dx\;\exp\{-(\alpha x^4+\beta
x^2+\gamma x)\}\
 \nonumber
\end {equation}
for $Re\: \alpha>0$.

Standard perturbative procedure rely on Taylor's decomposition of
$\exp(-\alpha x^4)$ term with consecutive replacements of the
integration and summation order. The integrals can be calculated,
but the sum is divergent.

We propose the power expansion in $\gamma$:
\begin {equation}
I_1=\sum\limits _{n=0} ^{\infty} \frac{(-\gamma)^n} {n!}\int\limits
_{-\infty}^{+\infty}\;dx\;x^n\exp\{-(\alpha x^4+\beta x^2)\}\ .
 \nonumber
\end {equation}

The integral is given in terms of the parabolic cylinder functions:

\begin {equation}
 D_{-m-1/2}(z)\ =\
\frac{e^{-z^2/4}}{\Gamma(m+1/2)}\int_0^\infty\;dx\;
 x^{m-1/2}\exp\{-\frac{1}{2}x^2-zx\}\ .
 \nonumber
\end {equation}
The integral $I_1$ then can be read:
\begin {equation}
 I_1=\frac{\Gamma(1/2)}{\sqrt{\beta}}\sum\limits _{m=0} ^{\infty}
 \frac{\xi^m}{m!}\mathcal{D}_{-m-1/2}(z)\ ,
\label{one1}
\end {equation}
where
$$\xi=\frac{\gamma^2}{4\beta}\ ,\
 z=\frac{\beta}{\sqrt{2\alpha}}\ ,\nonumber
$$
and we have used the abbreviation:
$$
 \mathcal{D}_{-m-1/2}(z) =  z^{m+1/2}
e^{\frac{\scriptstyle z^2}{\scriptstyle 4}}\; D_{-m-1/2}(z)\
.\nonumber
$$

\noindent
 It was shown, that sum in Eq.(\ref{one1}) is convergent
and for finite values of the parameters of the model this sum
converges uniformly.

Applying this idea of integration on the $N$ dimensional integral
(\ref{findim}) integral we have the result:
\begin {equation}
\mathcal{Z}_{N} = \left[\prod_{i=0}^{N}2(1+b\triangle^2/c)
\omega_i\right]^{-\frac{1}{2}} \; \mathcal{S}_{N} \nonumber
\end {equation}
with
\begin {equation}
\mathcal{S}_{N}= \sum\limits_{k_1,\cdots,k_{N-1}=0}^\infty \prod
\limits _{i=0}^N \; \left[
\frac{\left(\rho\right)^{2k_{i}}}{(2k_{i})!}
\Gamma(k_{i-1}+k_{i}+1/2)\; \sqrt{\omega_i}\;
\mathcal{D}_{-k_{i-1}-k_{i}-1/2}\;(z)\right], \label{rec1}
\end {equation}

\noindent where the constants and symbols in the above  relation are
connected to the constants of the model by the relations:  $k_0 =
k_N = 0,$ $\rho=(1+b\triangle^2/c)^{-1}$,
$z=c(1+b\triangle^2/c)/\sqrt{2a\triangle^3}$,
$\omega_i=1-A^2/\omega_{i-1}$, $\omega_0=1/2+b\triangle^2/c$,
$A=\frac{1}{2(1+b\triangle^2/c)}$.

 In the formula for $\mathcal{S}_{N}$ is useful to mention that:

- $\rho$ is independent of the coupling constant;

- only the argument $z$ of parabolic cylinder function is coupling
constant dependent;

It was shown, that for finite values of the parameters of the model
and one summation index $k_i\rightarrow\infty$ the $k_i - th$ term
of the sum approaches to zero as

$$\frac{k_i^{\alpha}\beta^{k_i}}{k_i!\; \exp{(\sqrt{k_i})}}\; ,$$
where $\alpha$ and $\beta$ are finite numbers. This asymptotic is
sufficient for a proof of the uniform convergence of the series for
$S_N$ not only for single $k_i$, but for arbitrary tuple $\{k_i\}$
of indices.

Following the idea of Gelfand and Yaglom the functional integral in
the continuum limit is defined by the formal limit
$$\lim_{N\rightarrow\infty}\, \mathcal{Z_N} = \frac{1}{\sqrt{F(\beta)}}\ ,$$
where  $F(\beta)$, for our case, is the solution of the differential
equation:
\begin {equation}
\frac{\partial^2}{\partial
\tau^2}F(\tau)+4\left(\frac{\partial}{\partial \tau}F(\tau)\right)
\; \left(\frac{\partial}{\partial
\tau}\ln{\mathcal{S}(\tau)}\right)=
F(\tau)\left(\frac{2b}{c}-2\frac{\partial^2}{\partial
\tau^2}\ln{\mathcal{S}(\tau)}-4(\frac{\partial}{\partial
\tau}\ln{\mathcal{S}(\tau)})^2\right)\ ,\ \tau\in (0, \beta)
\label{gy1}
\end {equation}

\noindent calculated at the point $\beta$ (the upper limit of the
time interval in the continuum action) with the initial conditions:
$F(0) = 1$ and $\partial F(\tau)/\partial \tau |_{\tau=o}= 0$.
Evaluating the continuum limit of the difference equation we use the
convention for definition of the continuum variable $\tau = n\
\triangle\ .$

The function $\mathcal{S}(\tau)$ is given as the continuum limit of
the Eq. (\ref{rec1})
$$\mathcal{S}(\tau) = \lim_{N \rightarrow \infty} \mathcal{S}_N\ .$$

Equation (\ref{gy1}) can be simplified by the substitution:
$$F(\tau) = \frac{y(\tau)}{\mathcal{S}(\tau)^2}\ .$$
For $y(\tau)$ we find the equation:
\begin {equation}
\frac{\partial^2}{\partial \tau^2}y(\tau) =
y(\tau)\left(\frac{2b}{c}  \right)\ , \label{gy2}
\end {equation}
accompanied  by the  initial conditions:

$$y(0) = \mathcal{S}(0)^2\ ,\
\left. \left. \frac{\partial y(\tau)}{\partial \tau}\right|_{\tau=0}
= \frac{\partial}{\partial \tau}\mathcal{S}(\tau)^2\right|_{\tau=0}.
$$
\noindent For harmonic oscillator we have $\mathcal{S}=1.$

To evaluate $S_N$, we must solve the problem how to sum up the
product of two parabolic cylinder functions in Eq. (\ref{rec1}). The
parabolic cylinder functions are the representation of the group of
the upper triangular matrices, so we implicitly expect the
simplification of the product due to a group principles. This
problem was not solved completely yet. We adopt less complex method
of summation, namely we use the asymptotic expansion one of
parabolic cylinder function, with precise sum over $k_i$ of the rest
of the relation, containing the other function. In the algebra such
relation for precise summation is available. Surely, the result is
degraded to the form of an asymptotic expansion only, but still we
shall have an analytical solution of the problem. This procedure is
widely discussed in article I, here we repeat the result:

\begin {eqnarray}
&&\mathcal{Z}_N\;=\;\left\{\prod\limits_{i=0}^{N}\;\left[2(1+b\triangle^2/c)\omega_
i\right]\right\}^{-1/2}\ \mathcal{S}_N\ , \label{rov11}\\
\noalign{\bigskip}
&&\mathcal{S}_{\Lambda}\ =
\sum\limits_{\mu=0}^{\mathcal{J}}\;\frac{(-1)^{\mu}}{\mu!\;(2z^2\triangle^3)^{\mu}}\;
\triangle^{3\mu}\left(\Lambda\right)^{2 \mu}_{0}, \label{rov12}
\end {eqnarray}

where the symbols $\left(\Lambda\right)^{2j}_{i}$ satisfy the
following recurrence relation:

\begin {equation}
\left(\Lambda\right)^{2\mu}_{2\mu-p}=
\sum\limits_{\lambda=0}^{\mu}\;\binom{\mu}{\lambda}\frac{1}{\omega_{\Lambda-1}^{2\mu-2\lambda}}
\sum\limits^{2\lambda}_{i=\max{[0,\; 2\lambda-p]}}
\left(\frac{A^2}{\omega_{\Lambda-2}\omega_{\Lambda-1}}\right)^i\;
\left(\Lambda-1\right)^{2\lambda}_{i}\; a_{2\mu-p}^{2\mu-2\lambda+i}
\label{recur1}
\end {equation}
The recurrence procedure begins from:
$$\left(1\right)^{2\lambda}_{i} = \frac{1}{\omega_0^{2\lambda}}a_i^{2\lambda}\ .$$
We repeat also the definition of the value $z,$ where the dependence on the coupling constant is hidden:
$$z=\frac{c(1+b\triangle^2/c)}{\sqrt{2a\triangle^3}}\ .$$

As follows in the calculation  the important role plays the objects
$\omega_i$ defined by recurrence as $$\omega_i\; =\;
1-\frac{A^2}{\omega_{i-1}}\ ,$$ with the first term for
unconditional measure integral:
$$\omega_0 \; =\; 1/2 + \frac{b\triangle^2/c}{2(1+b\triangle^2/c)}.$$

For the forthcoming calculation we will use the more convenient
variables, introduced in Appendix 2 of article I:
$$ Q_{i}= w_1 x^i +
w_2 y^i,$$ $$\tilde{Q}_{i}=w_1 x^i - w_2 y^i,$$ where
$$x=\frac{1}{2A}+\sqrt{\frac{1}{4A^2}-1},\; y=\frac{1}{2A}-\sqrt{\frac{1}{4A^2}-1},\; x y = 1,$$
$$w_1= 1 + \frac{2B}{\sqrt{1-4A^2}},\; w_2= 1 - \frac{2B}{\sqrt{1-4A^2}}.$$
$$A=\frac{1}{2(1+b\triangle^2/c)},\ B=\frac{b\triangle^2/c}{2(1+b\triangle^2/c)}.$$
We also use the fruitful identity following from the definition of
the $n-th$ convergent of the continued fraction, calculated in
article I:
$$ \frac{A}{\omega_{k-1}} = \frac{Q_{k-1}}{Q_k} $$

The symbols $a^j_i$ were defined by need to rewrite the Pochhammer
symbols $\pochh{k+1/2}{j}$ in the form:
$$\pochh{k+1/2}{j} = \sum^{\min{(j,k)}}_{i=0}a^j_i k(k-1) ... (k-i+1)\ ,$$
and by help the recurrence procedure we found:
$$a^j_i = \binom{j}{i}\frac{\pochh{1/2}{j}}{\pochh{1/2}{i}}\ .$$

In this article we explicitly evaluate the recurrence relation for
$\left(\Lambda\right)_{2\mu-p}^{2\mu}$ and we calculate  the
continuum limit of the function $\mathcal{S}_{\Lambda}$.

\section*{Evaluation of the recurrence relation}

We rewrite the recurrence relation (\ref{recur1}) into more
convenient form for consecutive calculation. We introduce the
quantities $Q_k$ by the identity:
$$ \frac{A}{\omega_{k-1}} = \frac{Q_{k-1}}{Q_k}\ . $$
We replace the summation index $i$ by the summation index $j$
defined by:
$$i=2\lambda - j\ .$$
Finally, we interchange the order of summations over indexes $j$ and
$\lambda$. We read:
\begin {eqnarray}
&&(A Q_{\Lambda} Q_{\Lambda-1})^{2d-p}
\left(\Lambda\right)_{2d-p}^{2d} = \nonumber
\\ \noalign{\bigskip}
&=&\sum\limits_{j=0}^p\;\frac{a^{2d-j}_{2d-p}}{(A Q_{\Lambda}
Q_{\Lambda-1})^{p-j}}\sum\limits^{d}_{\lambda=[\frac{j+1}{2}]}
\left(A Q_{\Lambda-2}Q_{\Lambda-1}\right)^{2\lambda-j}\;
\left(\Lambda-1\right)_{2\lambda-j}^{2\lambda}\;
\binom{d}{\lambda}(Q_{\Lambda-1}^4)^{d-\lambda}\ ,\ p\in <0,
2d>
\end {eqnarray}

The right hand side of the equation is $(2d,\ p)$-th matrix
element of the products of three matrices. For fixed $d$ and
 $p$, on the left hand side of the equation, we read only the
$d$ - th column of a matrix, which is recurrently tied to the
matrix in the center of the product on left hand side. We can use
the notation:

$$\ms X^d_{p,\mu}(\Lambda) = \sum\limits_{j=0}^p
\sum\limits^{\mu}_{\lambda=[\frac{j+1}{2}]}\; \ms
A^d_{p,j}(\Lambda-1)\ms C^d_{j,\lambda}(\Lambda-1)\ms
M^d_{\lambda,\mu}(\Lambda-1) .$$

The definition of the matrices $\ms A^d(\Lambda-1),\  \ms
C^d(\Lambda-1),\  \ms M^d(\Lambda-1)$ is the following:

1. The  $\ms A^d$ is the lower triangular matrix with the zeros over
the main diagonal of the dimension $(2\mu+1)(2\mu+1)$. The principal
minor of the dimension $(2d+1)(2d+1)$ is non-zero only with the
elements:
$$\left\{\ms A^d(\Lambda-1)\right\}_{p,j}= \frac{a^{2d-j}_{2d-p}}{(A Q_{\Lambda}
Q_{\Lambda-1})^{p-j}}\ .$$

2. The $\ms C_{j,\lambda}^d(\Lambda) $ is the upper triangular
matrix with the zeros under the main diagonal of the dimension
$(2\mu+1)(\mu+1)$. The nonzero elements form the main minor of the
dimension $(2d+1)(d+1)$ with $\lambda^{th}$ column:
$$\ms C_{p,\lambda}^d(\Lambda)=(A Q_{\Lambda} Q_{\Lambda-1})^{2\lambda-p}
\left(\Lambda\right)_{2\lambda-p}^{2\lambda},$$ where $p = 0, 1,
..., 2\lambda$ and $\lambda = 0, 1, ..., d\ $.

3. The $\ms M_{\lambda,k}^d$ is the upper triangular matrix with the
zeros under the main diagonal of the dimension $(\mu+1)(\mu+1)$. The
nonzero elements form the main minor of the dimension $(d+1)(d+1)\
$:
$$\ms M_{\lambda,k}^d(\Lambda-1) =
\binom{k}{\lambda}(Q^4_{\Lambda-1})^{k-\lambda} ,\  d\geq k\geq
\lambda \ge 0.$$

To evaluate the matrix $\ms C(\Lambda)$, we must calculate $\ms
X(\Lambda)$ for all dimensions up to $\mu$, for each dimension to
extract the last column of matrix $\ms X(\Lambda)$ and from these
columns to compose the matrix $\ms C(\Lambda)$. We define such
linear operation as follows:

1. Let $\ms A^d$ and $\ms M^d$ are the matrices of the dimensions
$(2\mu+1)(2\mu+1)$ and $(\mu+1)*\mu+1)$ respectively. Let $\ms C^d$
is the matrix of dimensions $(2\mu+1)(\mu+1)$. This matrices
possesses the nonzero main minors of the dimensions $(2d+1)(2d+1)$,
$(d+1)(d+1)$, and $(2d+1)(d+1)$ respectively.

2. Let $sup \ms M$ is the supermatrix possessing on the $d-th$ place
of the main diagonal the matrix $\ms M^d$ the same is defined for
$sup \ms A\ $.

3. The $sup \ms C$ is the supermatrix with $d-th$ diagonal element
of the form: $$\ms C^d (\Lambda) = \sum_{i_{\Lambda}=0}^d \tilde{\ms
X}^{d-i_{\Lambda}}(\Lambda)\ .$$

4. Matrix $\tilde{\ms X}^d$ is the one column matrix defined by the
relation:
$$\tilde{\ms X}^d(\Lambda) = \ms X^d(\Lambda)*\ms P^d,$$
where $\ms P^d$ is the projector of the d-th column of the matrix
$\ms X^d(\Lambda)$ into d-th column of the matrix $\tilde{\ms
X}^d(\Lambda)$. $\ms P^d$ is matrix with only nonzero term
$$\left\{\ms P^d\right\}_{d,k} = \delta_{d,k}.$$

5. The matrix $\ms X^d(\Lambda)$ is defined by relation:
$$\ms X^d(\Lambda) = \ms A^d(\Lambda-1)*\ms C^d(\Lambda-1)*\ms M^d(\Lambda-1).$$

6. Then, for $\ms C^d(\Lambda)$ we have the result:
$$\ms C^d(\Lambda) = \sum_{i_{\Lambda}=0}^d
\ms A^{d-i_{\Lambda}}(\Lambda-1)*\ms
C^{d-i_{\Lambda}}(\Lambda-1)*\tilde{\ms
M}^{d-i_{\Lambda}}(\Lambda-1).$$

7. After evaluation of the full recurrence we find:
\begin {eqnarray}
&&\ms C^d(\Lambda) = \sum_{i_{\Lambda}=0}^d
\sum_{i_{\Lambda-1}=0}^{d-i_{\Lambda}}\cdots
\sum_{i_2=0}^{d-i_{\Lambda}-i_{\Lambda-1}- \cdots -i_3}
\label{matr1}\\
\noalign{\bigskip}\nonumber &&\left\{\ms
A^{d-i_{\Lambda}}(\Lambda-1)*\ms
A^{d-i_{\Lambda}-i_{\Lambda-1}}(\Lambda-2)* \cdots
*\ms A^{d-i_{\Lambda}-i_{\Lambda-1}-\cdots-i_2}(1)\right\}*\ms C^{d-i_{\Lambda}-i_{\Lambda-1}-\cdots-i_2}(1)*
\\ \noalign{\bigskip}
\nonumber &&\left\{\tilde{\ms
M}^{d-i_{\Lambda}-i_{\Lambda-1}-\cdots-i_2}(1)*\cdots* \tilde{\ms
M}^{d-i_{\Lambda}-i_{\Lambda-1}}(\Lambda-2)* \tilde{\ms
M}^{d-i_{\Lambda}}(\Lambda-1)\right\}
\end {eqnarray}

To evaluate the product of two consecutive matrices from
multi-product:
$$\left\{\ms A^{d-i_{\Lambda}}(\Lambda-1)*\ms A^{d-i_{\Lambda}-i_{\Lambda-1}}(\Lambda-2)*
\cdots
*\ms A^{d-i_{\Lambda}-i_{\Lambda-1}-\cdots-i_2}(1)\right\}$$
we use the two identities for the summation over the index $j$:

$$a^{2I_3-j}_{2I_3-p}*a^{2I_2-\lambda}_{2I_2-j} =2^{-2(p-\lambda)}
\frac{(4
I_2-2\lambda)!}{(4I_3-2p)!(p-\lambda)!}\binom{p-\lambda}{j-\lambda}
\frac{(4I_3-2j)!}{(4I_2-2j)!}$$ and
$$\frac{(4I_3-2j)!}{(4I_2-2j)!} = \partial_\epsilon^{4I_3-4I_2}(\epsilon^{4I_3-2j})|_{\epsilon=1}$$

We introduced the abbreviation
 $$I_j = d - (i_j +i_{j+1} + \cdots +i_{\Lambda} )$$
 Then, for product of two lover-triangular matrices we have:

\begin {eqnarray}
\hspace{-2.0cm} &&\sum^p_{j=\lambda}\left\{\ms
A^{I_3}(2)\right\}_{p,j}\left\{\ms A^{I_2}(1)\right\}_{j,\lambda}=
\sum^p_{j=\lambda} \frac{a^{2I_3-j}_{2I_3-p}}{(AQ_3Q_2)^{p-j}}*
\frac{a^{2I_2-\lambda}_{2I_2-j}}{(AQ_2Q_1)^{j-\lambda}} =\\
\noalign{\bigskip}\nonumber
 &=&2^{-2(p-\lambda)} \frac{(4
I_2-2\lambda)!}{(4I_3-2p)!(p-\lambda)!}
\partial_\epsilon^{4i_2}\left\{(\epsilon^{4I_3-2p})
\sum^p_{j=\lambda}\binom{p-\lambda}{j-\lambda}\left(\frac{\epsilon^2}{AQ_3Q_2}\right)^{p-j}
\left(\frac{1}{AQ_2Q_1}\right)^{j-\lambda}\right\}\Bigg|_{\epsilon=1}
\end {eqnarray}

In the above relation the summation over index $j$ can be performed
explicitly. By the substitution:
$$\epsilon^2 = \xi ,$$

we find:

$$\sum^p_{j=\lambda}\left\{\ms A^{I_3}(2)\right\}_{p,j}\left\{\ms A^{I_2}(1)\right\}_{j,\lambda} =
2^{-2(p-\lambda)} \frac{(4 I_2-2\lambda)!}{(4I_3-2p)!(p-\lambda)!}
2^{4i_2}D^{i_2}_{\xi}\left\{\frac{1}{\xi^{p-2I_3}}
\left(\frac{\xi}{AQ_3Q_2}+
\frac{1}{AQ_2Q_1}\right)^{p-\lambda}\right\}\Bigg|_{\xi=1} \ ,$$

where $D_{\xi}$ is the differential operator calculated from
$\partial^4_{\epsilon}$ given as:

$$D_{\xi}=3/4\partial^2_{\xi}+3 \xi \partial^3_{\xi}+\xi^2 \partial^4_{\xi}$$

For  the resulting product of all matrices $\ms A^I(k)$ we find:

$$\left\{\ms A^{d-i_{\Lambda}}(\Lambda-1)*\ms A^{d-i_{\Lambda}-i_{\Lambda-1}}(\Lambda-2)*
\cdots
*\ms A^{d-i_{\Lambda}-i_{\Lambda-1}-\cdots-i_2}(1)\right\}_{p,\lambda}=$$
$$
= 2^{-2(p-\lambda)}
\frac{(4I_2-2\lambda)!}{(4I_{\Lambda}-2p)!(p-\lambda)!}
2^{4(i_2+i_3+\cdots+i_{\Lambda})}\times$$

$$\times\left\{\prod_{m=2}^{\Lambda}D^{i_m}_{\xi_m}
\left[\frac{1}{\xi_m^{p-2I_{m+1}}} \left(
\frac{1}{AQ_2Q_1}+\frac{\xi_2}{AQ_3Q_2}+\cdots+
\frac{\xi_2\cdots\xi_{\Lambda-1}}{AQ_{\Lambda}Q_{\Lambda-1}}
\right)^{p-\lambda} \right]\right\}\Bigg |_{(all \xi_m \rightarrow
1)}
$$

Evaluating the product of matrices:

$$\left\{\tilde{\ms M}^{d-i_{\Lambda}-i_{\Lambda-1}-\cdots-i_2}(1)*\cdots*
\tilde{\ms M}^{d-i_{\Lambda}-i_{\Lambda-1}}(\Lambda-2)* \tilde{\ms
M}^{d-i_{\Lambda}}(\Lambda-1)\right\}$$

we use that:

$\tilde{\ms M}^{I_{j+1}}(j)$ is one-column matrix with $I_{j+1}$
non-zero elements in $j+1$ column:

$$\left\{\tilde{\ms M}^{I_{j+1}}(j)\right\}_{\lambda,j+1} =
\binom{I_{j+1}}{\lambda}Q_j^{4(I_{j+1}-\lambda)},$$ where
$\lambda=0, 1, \cdots , I_{j+1}.$ Product of such matrices is
one-column matrix with the elements:

$$\left\{\tilde{\ms M}^{d-i_{\Lambda}-i_{\Lambda-1}-\cdots-i_2}(1)*\cdots*
\tilde{\ms M}^{d-i_{\Lambda}-i_{\Lambda-1}}(\Lambda-2)* \tilde{\ms
M}^{d-i_{\Lambda}}(\Lambda-1)\right\}_{\lambda,I_{\Lambda}}=$$

$$=\binom{I_2}{\lambda}\binom{I_3}{I_2} \cdots \binom{I_{\Lambda}}{I_{\Lambda-1}}
Q_1^{4(I_2-\lambda)}Q_2^{4(I_3-I_2)}\cdots
Q_{\Lambda-1}^{4(I_{\Lambda}-I_{\Lambda-1})}$$

From definition (\ref{recur1}) of the recurrence steps we have for
the matrix $\ms C^{I_2}(1)$ the nonzero elements:

$$\left\{\ms C^{I_2}(1)\right\}_{j,\lambda}=
\frac{Q_0^{4\lambda}}{(AQ_1Q_0)^j}a^{2\lambda}_{2\lambda-j}$$ with
the conditions for indices: $$0\leq j \leq 2\lambda \leq 2I_2 \leq
2\mu$$

 Collecting all partial results together, inserting them into Eq. (\ref{matr1}) and remember that for
function $\mathcal{S}_{\Lambda}$ defined in Eq. (\ref{rov12}) only
matrix elements $\left\{\ms C(\Lambda)^{2\mu}\right\}_{2\mu, 2\mu}$
are important, we find the result:

\begin {eqnarray}
&&\left\{\ms C(\Lambda)^{2\mu}\right\}_{2\mu, 2\mu} =
\sum_{i_{\Lambda}=0}^{\mu}\;
\sum_{i_{\Lambda-1}=0}^{\mu-i_{\Lambda}}\cdots
\sum_{i_2=0}^{\mu-i_{\Lambda}-i_{\Lambda-1}- \cdots -i_3}
\binom{I_3}{I_2} \cdots \binom{I_{\Lambda}}{I_{\Lambda-1}}
Q_2^{4(I_3-I_2)}\cdots
Q_{\Lambda-1}^{4(I_{\Lambda}-I_{\Lambda-1})}\times\nonumber
\\ \noalign{\bigskip}\nonumber
&&\times\sum_{j=0}^{2I_2}\sum_{\lambda =
\mid\frac{j+1}{2}\mid}^{I_2}
\frac{(4I_2-2j)!}{(2\mu-j)!}\frac{2^{4(i_2  + \cdots +
i_{\Lambda})}}{2^{(4\mu-2j)}}
\left\{\prod_{m=2}^{\Lambda}D^{i_m}_{\xi_m}
\left[\frac{1}{\xi_m^{2\mu-2I_{m+1}}} \left(
\frac{1}{AQ_2Q_1}+\cdots+
\frac{\xi_2\cdot\cdot\cdot\xi_{\Lambda-1}}{AQ_{\Lambda}Q_{\Lambda-1}}
\right)^{2\mu-j} \right]\right\}\Bigg|_{(all \xi_m\rightarrow1)}\times\\
\noalign{\bigskip}
 &&\times\left(\frac{1}{AQ_0Q_1}\right)^j
\left\{\binom{I_2}{\lambda}a^{2\lambda}_{2\lambda-j}\ Q^{4\lambda}_0
Q_1^{4(I_2-\lambda)}\right\} \label{maineq}
\end {eqnarray}

The asymptotic function $\mathcal{S}_{\Lambda}$ can be read:

\begin {equation}
\mathcal{S}_{\Lambda}=\sum\limits_{\mu=0}^{\mathcal{J}}\;\frac{(-1)^{\mu}}{\mu!\;(2z^2\triangle^3)^{\mu}}\;
\triangle^{3\mu}\left\{\ms C(\Lambda)^{2\mu}\right\}_{2\mu, 2\mu}\ ,
\label{maineq2}
\end {equation}
because $\{\ms C(\Lambda)^{2\mu}\}_{2\mu, 2\mu} =
(\Lambda)^{2\mu}_0$

Due to the analytic form for $\left\{\ms
C(\Lambda)^{2\mu}\right\}_{2\mu, 2\mu}$ we can express
$\mathcal{S}_{\Lambda}$ in the continuum $\triangle\rightarrow 0$
limit as well as in asymptotic $\mu\rightarrow \infty$ limit.

 \section*{The continuum limit}

In continuum limit we must take into account that:

-  in Eq. (\ref{maineq2}) the $\mu-th$ term $\left\{\ms
C(\Lambda)^{2\mu}\right\}_{2\mu,2\mu}$ is multiplied by
$\triangle^{3\mu}$.

- the sum $$\left( \frac{1}{AQ_2Q_1}+\frac{\xi_2}{AQ_3Q_2}+\cdots+
\frac{\xi_2\cdots\xi_{\Lambda-1}}{AQ_{\Lambda}Q_{\Lambda-1}}
\right)_{(all \xi_m\rightarrow1)}$$ is given by exact formula:
$$\frac{1}{2Aw_1w_2(x-y)}\left(\frac{\tilde{Q}_{\Lambda}}{Q_{\Lambda}} - \frac{\tilde{Q}_{0}}{Q_{0}}\right).$$
From the relation $$\frac{1}{2Aw_1w_2(x-y)}\sim
\frac{1}{\sqrt{2b/c}\ \triangle},$$ we deduce that to the leading
term of Eq. (\ref{maineq}) contribute the terms with summation index
$j=0$ with the contribution proportional to
$\left(\frac{1}{\triangle}\right)^{2\mu}$.

- to obtain the additional necessary factor
$\left(\frac{1}{\triangle}\right)^{\mu}$ the leading term must be
composed from the contributions where $\mu$ summation indices of the
$\left\{i_2, i_3, \cdots , i_{\Lambda-1}\right\}$ are equal to $1$.
Let it be the combination, say, $\{i_{n_1}, i_{n_2}, \cdots ,
i_{n_{\mu}}\}$. The sum of these indices is equal to $\mu$, and in
such case also $I_2=0$, because of definition $I_2=\mu-i_{n_1}-
i_{n_2}- \cdots -i_{n_{\mu}}$. The contribution to the body of
principal formula (\ref{maineq}) then can be read:

\begin {eqnarray}
&& \binom{I_{n_2}}{I_{n_1}}\binom{I_{n_3}}{I_{n_2}}\cdots
\binom{I_{n_{\mu}}}{I_{n_{\mu-1}}}
 Q_{n_1}^{4}\cdots Q_{n_{\mu}}^{4}
\\ \noalign{\bigskip}\nonumber
&& \frac{1}{(2\mu)!} \left\{\prod_{m=1}^{\mu}D_{\xi_{n_m}}
\frac{1}{\xi_{n_m}^{2\mu-2I_{n_{m+1}}}}\right\}\left. \left(
\frac{1}{AQ_2Q_1}+\frac{\xi_2}{AQ_3Q_2}+\cdot\cdot\cdot+
\frac{\xi_2\cdot\cdot\cdot\xi_{\Lambda-1}}{AQ_{\Lambda}Q_{\Lambda-1}}
\right)^{2\mu} \right]_{(all \xi\rightarrow1)}
\end {eqnarray}

 We must sum over all
such combinations, this can be done by $\mu$ summations. Every
summation is proportional to $\triangle^{-1}$. Therefore leading
term proportional to $\triangle^{-3\mu}$ can be achieved only if
$\mu$ different indices $i_n = 1$. Taking into account, that the
difference of two consecutive $I_{n_i}$ is one, we can rewrite the
dominant contribution in the continuum limit into the form:

\begin {eqnarray}
&&\left\{\ms C(\Lambda)^{2\mu}\right\}_{2\mu,2\mu} =
\sum_{n_{1}=2}^{\Lambda-\mu}\ \sum_{n_2=n_1+1}^{\Lambda-\mu+1}\cdots
\sum_{n_{\mu}=n_{\mu-1}+1}^{\Lambda}\mu! Q_{n_1}^{4}\cdots
Q_{n_{\mu}}^{4}
\\ \noalign{\bigskip}\nonumber
&& \frac{1}{(2\mu)!} \left\{\prod_{m=1}^{\mu}D_{\xi_{n_m}}
\frac{1}{\xi_{n_m}^{2\mu-2I_{n_{m+1}}}}\right\}\left. \left(
\frac{1}{AQ_2Q_1}+\frac{\xi_2}{AQ_3Q_2}+\cdot\cdot\cdot+
\frac{\xi_2\cdot\cdot\cdot\xi_{\Lambda-1}}{AQ_{\Lambda}Q_{\Lambda-1}}
\right)^{2\mu} \right]_{(all \xi\rightarrow1)} \label{main}
\end {eqnarray}

\subsection*{The effect of the operator $D_{\xi}$.}

Let us evaluate the operation of the operator $D_{x}$ accompanied by
operator's variable term:

\begin {equation}
\lim_{x_j\rightarrow 1}D_{x_j}\ \frac{1}{x_j^{2(i_{j+1}+ \cdots +
i_{\Lambda})}}
\label{opr1}
\end {equation}

$i_j$ times on the $n-th$ power of the function

$$f_j = a_{j-1} + (x_2 \cdots x_j) b_j$$
where:

$$ a_{j-1} = \frac{1}{Q_2Q_1}+\frac{x_2}{Q_3Q_2}+ \cdots + \frac{x_2 \cdots x_{j-1}}{Q_jQ_{j-1}}$$

$$b_j = \frac{1}{Q_{j+1}Q_j}+ \cdots + \frac{1}{Q_{\Lambda}Q_{\Lambda}-1}.$$
This definition reflect the fact, that before application of
$D_{x_j}$ we provide the limits $x_{j+1}\rightarrow 1, \cdots ,
x_{\Lambda}\rightarrow 1.$ The following limits will be used:

$$\lim_{x_j\rightarrow 1}(x_2 \cdots x_j) = (x_2 \cdots x_{j-1})$$

$$\lim_{x_j\rightarrow 1}f_j = f_{j-1} = a_{j-2} + (x_2 \cdots x_{j-1})b_{j-1}$$

We meet very important feature of the application of the operator
$D_{x}$.

\noindent
1. Applying $D_{x}$ the first time for first nonzero
$i_j$, all $i_{j+1}, \cdots , i_{\Lambda}$ are zero, then factor
$\frac{1}{x_j^{2(i_{j+1}+ \cdots + i_{\Lambda})}} = 1.$

\noindent 2.
We find  in the evaluation,  that there always  appears
the terms, killing the variables in denominator of (\ref{opr1}) in
the next steps of calculation.

From the practical reasons, our aim is to express the resulting
formula in the form where the dependence on the next derivative
variable, $x_{j-1}$ is in the function $f_{j-1}$ only. We find:

\begin {eqnarray}
&&\lim_{x_j\rightarrow 1}D^{i_j}_{x_j}\ \frac{1}{x_j^{2(i_{j+1}+
\cdots + i_{\Lambda})}}f_j^n = \\ \noalign{\bigskip}\nonumber
&&[(x_2\cdots x_{j-1})b_j]^{2i_j}\sum_{l=0}^{MIN}\
\binom{2i_j}{l}(a_{j-2})^l\
(f_{j-1})^{n-2i_j-l}\textbf{J}(l,MIN;n,i_j;b_j/b_{j-1})
\end {eqnarray}

where
$$\textbf{J}(l,MIN;n,i_j;b_j/b_{j-1}) = \sum_{p=l}^{MIN}\ \binom{2i_j-l}{p-l}\ \binom{2i_j-1/2}{2i_j-p}\
\frac{n!(2i_j-p)!}{(n-2i_j-p)!}\ \left(\frac{b_j}{b_{j-1}}\right)^p
,$$

$$MIN = \min{(2i_j, n-2i_j)}.$$
The proof of this formula is given in Appendix A.

We see, that in the term $[(x_2\cdots x_{j-1})b_j]^{2i_j}$ is the
variable power $x^{2i_j}_k$ canceling the same power of the variable
in the denominator of the operator effecting over variable $x_k$, to
left the evaluation simpler. It can be shown, that function
$\textbf{J}$ is proportional to the Gegenbauer orthogonal polynomial
following the relation for the Jacobi orthogonal polynomial
$P_n^{(\alpha,\beta)}$ (see e.g. Prudnikov \cite{prud}):

\begin {eqnarray}
\sum_k^n\; \binom{n+\alpha}{k}\; \binom{n+\beta}{n-k}\;
\left(\frac{x+1}{x-1}\right)^k\; =\; \frac{2^n}{(x-1)^n}\;
P_n^{(\alpha,\beta)}(x)
\end {eqnarray}
If the $\beta$ is a half number, the Jacobi polynomial can be
expressed by help of the Gegenbauer polynomial.

The result of two consecutive application of the operator $D_{x_j}$
and $D_{x_{j-1}}$ is:

\begin {eqnarray}
&&\left\{\lim_{x_{j-1}\rightarrow 1}D^{i_{j-1}}_{x_{j-1}}\
\frac{1}{x_{j-1}^{2(i_{j}+ \cdots + i_{\Lambda})}}\right\}
\left\{\lim_{x_j\rightarrow 1}D^{i_j}_{x_j}\
\frac{1}{x_j^{2(i_{j+1}+ \cdots + i_{\Lambda})}}\right\}(f_j)^n=\\
\noalign{\bigskip}\nonumber &&[(x_2\cdots
x_{j-2})b_{j-1}]^{2i_{j-1}}[(x_2\cdots
x_{j-2})b_j]^{2i_j}\sum_{l=0}^{MIN1}\binom{2i_j}{l}(a_{j-2})^l
\sum_{\acute{l}=0}^{MIN2}\binom{2i_{j-1}}{\acute{l}}(a_{j-3})^{\acute{l}}\\
\noalign{\bigskip}\nonumber
&&(f_{j-2})^{n-2i_j-2i_{j-1}-l-\acute{l}}\
\textbf{J}(l,MIN1;n,i_j;b_j/b_{j-1})\
\textbf{J}(\acute{l},MIN2;n-2i_j-l,i_{j-1};b_{j-1}/b_{j-2})
\end {eqnarray}
where $$MIN1=\min{(2i_j, n-2i_j)},$$ $$MIN2=\min{(2i_{j-1},
n-2i_j-2i_{j-1}-l)}.$$

On the result of application of three operations the nonlinear
character of the our result is clearly visible:

\begin {eqnarray}
&&\left\{\lim_{x_{j-2}\rightarrow 1}D^{i_{j-2}}_{x_{j-2}}\
\frac{1}{x_{j-2}^{2(i_{j-1}+ \cdots + i_{\Lambda})}}\right\}
\left\{\lim_{x_{j-1}\rightarrow 1}D^{i_{j-1}}_{x_{j-1}}\
\frac{1}{x_{j-1}^{2(i_{j}+ \cdots + i_{\Lambda})}}\right\}
\left\{\lim_{x_j\rightarrow 1}D^{i_j}_{x_j}\
\frac{1}{x_j^{2(i_{j+1}+ \cdots + i_{\Lambda})}}\right\}(f_j)^n=\\
\noalign{\bigskip}\nonumber
&&[(x_2\cdots
x_{j-3})b_{j-2}]^{2i_{j-2}} [(x_2\cdots
x_{j-3})b_{j-1}]^{2i_{j-1}}[(x_2\cdots x_{j-3})b_j]^{2i_j}\\
\noalign{\bigskip}\nonumber
&&\sum_{m=0}^{MIN1}\binom{2i_{j}}{m}\left(\frac{b_{j-1}}{b_{j-2}}\right)^m\
\sum_{l=m}^{MIN1}\binom{2i_{j}-m}{l-m}\left(1-\frac{b_{j-1}}{b_{j-2}}\right)^{l-m}\
\sum_{\acute{l}=0}^{MIN2}\binom{2i_{j-1}}{\acute{l}}(a_{j-3})^{\acute{l}+m}\\
\noalign{\bigskip}\nonumber &&\sum_{\nu=0}^{MIN3}
\binom{2i_{j-2}}{\nu}(a_{j-4})^{\nu}(f_{j-3})^{n-2i_j-2i_{j-1}2i_{j-2}-\acute{l}-m-\nu}\
\textbf{J}(l,MIN1;n,i_j;b_j/b_{j-1})\ \\ \noalign{\bigskip}\nonumber
&&\textbf{J}(\acute{l},MIN2;n-2i_j-l,i_{j-1};b_{j-1}/b_{j-2})
\textbf{J}(\nu,MIN3;n-2i_j-2i_{j-1}-m-\acute{l},i_{j-2};b_{j-2}/b_{j-3})
\end {eqnarray}

where

$$MIN3 = \min{(2i_{j-2},n-2i_j-2i_{j-1}-2i_{j-2}-m-\acute{l})}.$$

The above relations give first three terms of the asymptotic
expansion for function $S_{\Lambda}$ (\ref{maineq2}). We find:

For $\mu=1$ only one summation index $i_j$ is nonzero, then
$$\left\{\ms C(\Lambda)^2\right\}_{2,2}=\frac{\mu!}{(2\mu)!}\sum_{k=2}^{\Lambda}\
Q_k^4\ b_k^2\ \textbf{J}(0,0;2,1;b_k/b_{k-1})$$

For $\mu=2$ two summation index $i_j$ are nonzero, then

$$\left\{\ms C(\Lambda)^4\right\}_{4,4}=\frac{\mu!}{(2\mu)!}\sum_{k=2}^{\Lambda-1}\sum_{p=k+1}^{\Lambda}\
Q_k^4Q_p^4\ b_k^2b_p^2\
\textbf{J}(0,2;4,1;b_k/b_{p})\textbf{J}(0,0;2,1;b_p/b_{p-1})$$

For $\mu=3$ three summation index $i_j$ are nonzero, then

\begin {eqnarray}
&&\left\{\ms C(\Lambda)^6\right\}_{6,6}=
\frac{\mu!}{(2\mu)!}\sum_{k=2}^{\Lambda-2}\sum_{p=k+1}^{\Lambda-1}\sum_{q=p+1}^{\Lambda}\
Q_k^4Q_p^4Q_q^4\ b_k^2b_p^2b_q^2\\ \noalign{\bigskip}\nonumber
&&\sum_{l=0}^2\binom{2}{l}\ \left(1-\frac{b_p}{b_q}\right)^l\
\textbf{J}(l,2;6,1;b_k/b_{p})\ \textbf{J}(0,2-l;4-l,1;b_p/b_{q})\
\textbf{J}(0,0;2,1;b_q/b_{q-1})
\end{eqnarray}

The continuum limit is introduced by prescription:

$$\triangle k \rightarrow x,\ \triangle\; \Lambda \rightarrow \tau,$$

$$\sum_{k=2}^{\Lambda} \rightarrow \frac{1}{\triangle}\
\int_0^{\tau}\ d x,$$ When $\Lambda \rightarrow N,$ then $\tau
\rightarrow \beta,$ where $\beta$ is the the constant of the model,
and $N= \frac{\beta}{\triangle}.$

In continuum limit we obtain:
$$Q_k \rightarrow 2\cosh(\gamma x)$$
$$b_k \rightarrow \frac{1}{\triangle \gamma}(\tanh(\gamma \tau)-\tanh(\gamma x))$$
where $\gamma = \sqrt{2b/c},$ $b$ and $c$ are the parameters of the
model. The continuum limit of the relation (\ref{maineq2}) we will
call $S(a,b,c,\tau)$. To ilustrate the analytical form of the
result, we show the first nontrivial term $(\mu=1):$

\begin {equation}
\left\{\ms C^2(a,b,c,\tau)\right\}_{2,2} = \frac{3}{8
\gamma^3}\left[3\gamma \tau \tanh^2(\gamma \tau) +
 \tanh(\gamma \tau) - \gamma \tau\right]
\end {equation}
 For the higher we have the analytical formulas also
as the results of algebraic evaluation by Mathematica. The continuum
function $S(a,b,c,\tau)$ for the first three nontrivial
contributions is shown in the Fig. 1.

\begin{figure}
  \includegraphics[width=9cm]{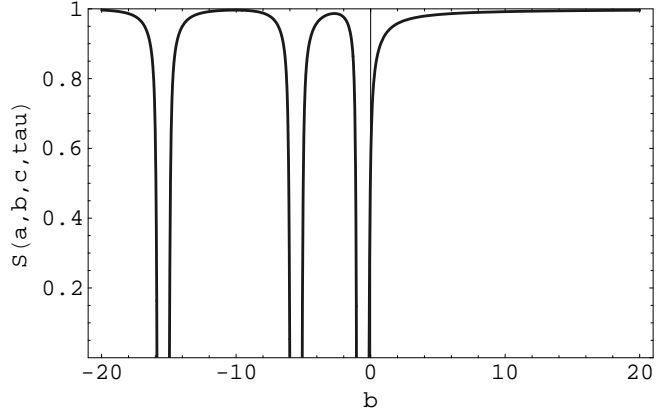}
  \caption {$b$ dependence of the continuum function $S(a,b,c,\tau)$ for fixed values $a, c, \tau.$ The first three nontrivial terms
 of the asymptotic series (\ref{maineq2}) were used.}
\end{figure}

The corresponding term for the Gelfand-Yaglom equation,
$-2\partial^2_{\mu}\ln(S(a,b,c,\tau))-4(\partial_{\mu}\ln(S(a,b,c,\tau)))^2$
is shown in the Fig. 2.

\begin{figure}
  \includegraphics[width=9cm]{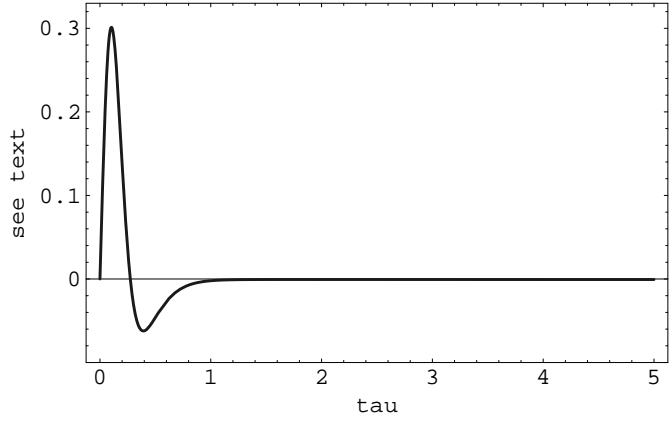}
 \caption{$\tau$ dependence of the continuum function
 $-2\partial^2_{\mu}\ln(S(a,b,c,\tau))-4(\partial_{\mu}\ln(S(a,b,c,\tau)))^2$ for fixed $a, b, c.$
 The first three nontrivial terms
 of the asymptotic series (\ref{maineq2}) were used.}
\end{figure}

 \section*{The leading divergent term in the limit $\mu \rightarrow \infty$}

 In this limit the terms (\ref{maineq}) are divergent. We
 are going to evaluate the leading divergent term for each $\mu$. To provide this,
 we must to evaluate the double sums in Eq. (\ref{maineq}):

\begin {eqnarray}
&&\sum_{j=0}^{2I_2}\sum_{\lambda = \mid\frac{j+1}{2}\mid}^{I_2}
\frac{(4I_2-2j)!}{(2\mu-j)!}\ \frac{2^{4(i_2  + \cdots +
i_{\Lambda})}}{2^{(4\mu-2j)}}
\left\{\prod_{m=2}^{\Lambda}D^{i_m}_{\xi_m}
\left[\frac{1}{\xi_m^{2\mu-2I_{m+1}}} \left(
\frac{1}{AQ_2Q_1}+\cdots+
\frac{\xi_2\cdot\cdot\cdot\xi_{\Lambda-1}}{AQ_{\Lambda}Q_{\Lambda-1}}
\right)^{2\mu-j} \right]\right\}_{(\xi_m\rightarrow1)}\nonumber \\
\noalign{\bigskip}
 &&\left(\frac{1}{AQ_0Q_1}\right)^j
\left\{\binom{I_2}{\lambda}a^{2\lambda}_{2\lambda-j}\ Q^{4\lambda}_0
Q_1^{4(I_2-\lambda)}\right\}
\end {eqnarray}

The details of the calculation are explained in the Appendix B, we
obtained the result:

$$\left\{\ms C(\Lambda)^{2\mu}\right\}_{2\mu,2\mu}\; =\; \pochh{1/2}{2\mu}\left( \frac{1}{AQ_2Q_1}+\cdots+
\frac{1}{AQ_{\Lambda}Q_{\Lambda-1}} \right)^{2\mu}
(Q_0^4+Q_1^4)^{\mu}\ \sim \ 2^{\mu}\
\pochh{1/2}{2\mu}\left(\frac{\tanh{(\gamma
\tau)}}{\triangle\gamma}\right)^{2\mu}$$

For the leading divergent term  $\mu \rightarrow \infty$  of the
asymptotic series for $\mathcal{S}_{\Lambda}$ (\ref{maineq2}) we
finally have:

\begin {equation}
\frac{(-1)^{\mu}a^{\mu}}{\mu!\;c^{2\mu}}\; \triangle^{\mu}2^{\mu}\
\pochh{1/2}{2\mu}\left(\frac{\tanh{(\gamma
\tau)}}{\gamma}\right)^{2\mu}
\end {equation}

The series of this form is an asymptotic expansion of the parabolic
cylinder function of the index $-1/2$ and the argument
$$ z^{-1} =2 a \triangle\left(\frac{\tanh{(\gamma
\tau)}}{c \gamma}\right)^2\ .$$

\section*{Gelfand - Yaglom equation for anharmonic oscillator with mass zero}

As a test of our calculation we evaluate the energy levels of
anharmonic oscillator with zero mass. The  Gelfand - Yaglom equation
for this case will be obtained by limit $b\rightarrow 0$, and we
have:

\begin {equation}
y''(\tau) + 4 g_0(\tau) y'(\tau) =  f_0(\tau) y(\tau)
\end {equation}

For the functions $f_0(\tau)$ and $g_0(\tau)$ we evaluate from Eq.
(\ref{maineq2}) for first three non-trivial terms:

\begin {eqnarray}
f_0(\tau)&=& -2\left[ \frac{d^2}{d \tau^2}\ln{\left(1-\frac{a
\tau^3}{c^2} + \frac{43}{30}\left(\frac{a \tau^3}{c^2}\right)^2 -
\frac{8111}{1890}\left(\frac{a \tau^3}{c^2}\right)^3 + \cdots
\right)} \right]\\
&-&4\left[ \frac{d}{d \tau}\ln{\left(1-\frac{a \tau^3}{c^2} +
\frac{43}{30}\left(\frac{a \tau^3}{c^2}\right)^2 -
\frac{8111}{1890}\left(\frac{a \tau^3}{c^2}\right)^3 + \cdots
\right)} \right]^2\ ,\nonumber \\
g_0(\tau)&=& \left[ \frac{d}{d \tau}\ln{\left(1-\frac{a \tau^3}{c^2}
+ \frac{43}{30}\left(\frac{a \tau^3}{c^2}\right)^2 -
\frac{8111}{1890}\left(\frac{a \tau^3}{c^2}\right)^3 + \cdots
\right)} \right]\ .\nonumber
\end {eqnarray}

For first order in the coupling constant $a$, we have the equation:

\begin {equation}
y''(\tau)- \frac{3.4 a \tau^2}{c^2}y'(\tau) = \frac{3.4 a \tau}{c^2}
y(\tau)
\end {equation}

This equation can be solved analytically, (see Kamke \cite{kam}, Eq.
2.60). By substitution $y(\tau)=u(\tau)\exp{(2a\tau^3/c^2)}$ we find
the equation: $$u''(\tau)-\left(\frac{6a}{c^2}\right)^2\tau^4\
u(\tau)\ .$$ The general solution is expressed as the linear
combination of the Bessel functions:

\begin {equation}
y(\tau) = \sqrt{\tau}\left(C_1 J_{1/6}(i\frac{2a\tau^3}{c^2}) + C_2
Y_{1/6}(i\frac{2a\tau^3}{c^2})\right)\exp{\left(\frac{2a\tau^3}{c^2}\right)}\
.
\end {equation}
The constants $C_1$ and $C_2$ will be fixed from boundary
conditions. For $z\sim 0$ we follow the identities \cite{bateman}:
$$J_{\nu}(z)=\left(\frac{z}{2}\right)^{\nu}\sum_{m=0}^{\infty}
\frac{(-1)^m}{m!\ \Gamma(m+\nu+1)}\left(\frac{z}{2}\right)^{2m}\ ,$$
and $$Y_{\nu}(z)= \frac{1}{\sin(\nu \pi)}\left[J_{\nu}(z)\ \cos(\nu
\pi) - J_{-\nu}(z) \right]\ .$$ Inserting to the equations $y(0)=1$
and $y'(0)=0$ we find:
$$C_1 = -\left(\frac{ai}{c^2}\right)^{1/6}\Gamma(1-1/6)\sin{(\pi/6)}$$
$$C_2 = \left(\frac{ai}{c^2}\right)^{1/6}\Gamma(1-1/6)\cos{(\pi/6)}$$

For evaluation of the energy of the ground state we need $y(\tau)$
for $\tau \rightarrow \infty\ .$ For this limit we can use the
relations \cite{bateman}:
$$J_{\nu}(z)=\sqrt{\frac{2}{\pi z}}\cos\left(z-\frac{2\nu+1}{4}\pi\right)\ ,$$
$$Y_{\nu}(z)=\sqrt{\frac{2}{\pi z}}\sin\left(z-\frac{2\nu+1}{4}\pi\right)\ ,$$
and finally we have for the functional integral in this limit:
$$F(\beta)=\left[\left(\frac{ai}{c^2}\right)^{1/6}\Gamma(1-1/6)
\left(\frac{c^2}{ia\pi
\beta^3}\right)^{1/2}\sin\left(\frac{2ia\beta^3}{c^2}
-\frac{\pi}{6}\right)\exp{\left(\frac{2a\beta^3}{c^2}\right)}\right]^{-1/2}$$

The unconditional measure functional integral is the partition
function for the model solved. for the harmonic oscillator there is
the simple relation for the energy of the ground state:
$$ E_0 = -\lim_{\beta\rightarrow \infty}\left(\frac{1}{\beta}\ln{F(\beta)}\right)\ .$$
By the direct application of this relation we obtain zero for the
ground state energy of the an-harmonic oscillator with zero mass. We
find the nonzero result for the definition:
$$ E_0 = -\lim_{\beta\rightarrow \infty}\left(\frac{1}{\beta^3}\ln{F(\beta)}\right)\ = \frac{2a}{c^2}\ .$$

\section*{Conclusions}

In this article we calculated step-by-step the correlation function
for differential Gelfand-Yaglom equation an-harmonic oscillator with
positive or negative mass squared term. We stress, that generalized
Gelfand-Yaglom equation is the non-perturbative equation, the
correlation function is evaluated in the form of the asymptotic
series. The analytical form of the correlation function enables us
to evaluate the continuum limit as well as asymptotic limit. The
continuum limit can be used for evaluation such physical quantities
as energy levels. The asymptotic limit cam be used for attempts to
sum the series by Borel's method.

\vskip 1.3cm {\bf{Acknowledgements}}. This work was supported by
VEGA projects No. 2/6074/26.

\vskip 1.3cm
\appendix*{Appendix A}
\vskip 1.3cm

We are going to apply the derivative operator
\begin {equation}
D_{x_j}\ =3/2\partial^2_{x_j}+3 x_j \partial^3_{x_j}+x_j^2
\partial^4_{x_j}
\label{opra1}
\end {equation}

$i$ times on the $n-th$ power of the function

$$f_j = a_{j-1} + (x_2 \cdots x_j) b_j$$
where:

$$ a_{j-1} = \frac{1}{Q_2Q_1}+\frac{x_2}{Q_3Q_2}+ \cdots + \frac{x_2 \cdots x_{j-1}}{Q_jQ_{j-1}}$$

$$b_j = \frac{1}{Q_{j+1}Q_j}+ \cdots + \frac{1}{Q_{\Lambda}Q_{\Lambda}-1}$$
 We see, that $b_j$ is a constant from point of derivative operator
 $D_{x_j}$ and $a_{j-1}$ also. Therefore for $(D_{x_j})^i\ f_j^n$
  we find:

\begin {eqnarray}
& &(D_{x_j})^i\ f_j^n = \nonumber \\
& &((x_2 x_3 \cdots x_{j-1})b_j)^{2i} \sum_{m=0}^{min(2i, n-2i)}\
(-1)^m \binom{2i}{m}\ a^m_{j-1}\
\frac{n!}{(n-2i-m)!}\pochh{n-2i+1/2}{2i-m}\ f_j^{n-2i-m} \label{ap1}
\end {eqnarray}
The above result is nonzero only if $2i \leq n$. The variable
$x_{j-1}$ is in the term $f_j$ and $a_{j-1}$ as well. By the
identity $$a_{j-1}=f_j-(x_2 x_3 \cdots x_{j-1})b_j)$$ we introduce
summation due to binomial expansion of $a_{j-1}$  into
Eq.(\ref{ap1}) over index $p=0, 1, \cdots, m$. We exchange the order
of summations and apply the identity:
$$\binom{2i}{m} \binom{m}{p}
= \binom{2i}{p} \binom{2i-p}{m-p}$$

Finally, performing the limit $x_j \rightarrow 1$ we find:
\begin {eqnarray}
& &(D_{x_j})^i\ f_j^n|_{x_j=1} =((x_2 x_3 \cdots x_{j-1})b_j)^{2i}
\sum_{p=0}^{min(2i, n-2i)}\binom{2i}{p}((x_2 x_3 \cdots x_{j-1})b_j))^p f_{j-1}^{n-2i-p}\nonumber \\
& &\sum_{m=p}^{min(2i, n-2i)}\ (-1)^{p-m} \binom{2i-p}{m-p}\
\frac{n!}{(n-2i-m)!}\pochh{n-2i+1/2}{2i-m}
\label{ap2}
\end {eqnarray}
Now, we can perform the summation over index $m$. For the case
$min(2i, n-2i) = 2i$ we use the identities:

\begin {equation}
\frac{n!}{(n-2i-m)!} = \lim_{x\rightarrow 1}\partial_x^{2i+m}\left(
x^n \right)
\end {equation}
and
\begin {equation}
\pochh{n-2i+1/2}{2i-m} = \lim_{x\rightarrow
1}\partial_x^{2i-m}\left(\frac{(-1)^{2i-m}}{x^{n-2i+1/2}}\right)
\end {equation}
Inserting this into sum over $m$ in Eq (\ref{ap2}), replacing the
order of limit and summation, as well as the summation index $m$ by
$l=m-p$ we have:
\begin {eqnarray}
\lim_{x\rightarrow 1}\sum_{l=0}^{2i-p}\ (-1)^l\ \binom{2i-p}{l}\
\partial_x^l\left(\partial_x^{2i+p}\left( x^n \right)\right)
\partial_x^{2i-p-l}\left(\frac{(-1)^{2i-p-l}}{x^{n-2i+1/2}}\right)=
\frac{n!}{(n-2i-p)!}\pochh{p+1/2}{2i-p}
\end {eqnarray}
When $min(2i, n-2i) = n-2i$ we find the identical result. We proved
that:
\begin {eqnarray}
& &(D_{x_j})^i\ f_j^n|_{x_j=1} =\nonumber \\
& &((x_2 x_3 \cdots x_{j-1})b_j)^{2i} \sum_{p=0}^{min(2i,
n-2i)}\binom{2i}{p}((x_2 x_3 \cdots x_{j-1})b_j))^p
f_{j-1}^{n-2i-p}\ \frac{n!}{(n-2i-p)!}\pochh{p+1/2}{2i-p}
\label{ap3}
\end {eqnarray}
In the sum over index $p$ we recognize the formula for the Jacobi
orthogonal polynomial. By help of the identity:
$$\binom{2i}{p}\ \frac{n!}{(n-2i-p)!}\pochh{p+1/2}{2i-p} = \frac{(2i)!n!}{(n-2i)!}
\binom{n-2i}{p}\binom{2i-1/2}{2i-p} $$ for the case $min(2i, n-2i) =
2i$ we find the similarity with identity (see e.g.Prudnikov
\cite{prud}):

\begin {eqnarray}
\sum_{k=0}^n\; \binom{n+\alpha}{k}\; \binom{n+\beta}{n-k}\;
\left(\frac{x+1}{x-1}\right)^k\; =\; \frac{2^n}{(x-1)^n}\;
P_n^{(\alpha,\beta)}(x)
\end {eqnarray}
Where $P_n^{(\alpha,\beta)}(x)$ is Jacobi orthogonal polynomial.

The expression in Eq. (\ref{ap3}) we simplify further for
application of the operator $(D_{x_{j-1}}).$ In Eq. (\ref{ap3}) the
variable $x_{j-1}$ is in two terms, therefore we simplify that
equation by the identities:
$$(x_2 x_3 \cdots x_{j-1})b_j = \frac{b_j}{b_{j-1}}\ (x_2 x_3 \cdots x_{j-1})b_{j-1},$$
and
$$(x_2 x_3 \cdots x_{j-1})b_{j-1} = f_{j-1} - a_{j-2}.$$
Now, the variable $x_{j-1}$ will be in $f_{j-1}$ only and for
application of the operator $(D_{x_{j-1}})$ we prepare the relation:

\begin {eqnarray}
& &(D_{x_j})^i\ f_j^n \big|_{x_j=1} =\nonumber \\
& &((x_2 x_3 \cdots x_{j-1})b_j)^{2i} \sum_{p=0}^{min(2i,
n-2i)}\binom{2i}{p}\left(\frac{b_j}{b_{j-1}}\right)^p
\sum^p_{\nu=0}\binom{p}{\nu}\ (-a_{j-2})^{\nu} f_{j-1}^{n-2i-\nu}\
\frac{n!\ \pochh{p+1/2}{2i-p}}{(n-2i-p)!} \nonumber
\end {eqnarray}

By exchange of the order of summation we finally red:

\begin {eqnarray}
& &(D_{x_j})^i\ f_j^n \big|_{x_j=1} =\nonumber \\
& &((x_2 x_3 \cdots x_{j-1})b_j)^{2i} \sum^{min(2i,
n-2i)}_{\nu=0}\binom{2i}{\nu}\ (-a_{j-2})^{\nu} f_{j-1}^{n-2i-\nu}\
\sum_{p=\nu}^{min(2i,
n-2i)}\binom{2i-\nu}{p-\nu}\left(\frac{b_j}{b_{j-1}}\right)^p
\frac{n!\ \pochh{p+1/2}{2i-p}}{(n-2i-p)!} \nonumber
\end {eqnarray}
In this form is result of application of operator $(D_{x_j})^i$
suitable for the next evaluations.

\vspace{1.0cm}
\appendix*{Appendix B}
\vspace{1.0cm}

We are going to evaluate the leading divergent term. To provide
this,
 we must to evaluate the double sum in Eq. (\ref{maineq}):

\begin {eqnarray}
&&\sum_{j=0}^{2I_2}\sum_{\lambda = \mid\frac{j+1}{2}\mid}^{I_2}
\frac{(4I_2-2j)!}{(2\mu-j)!}\ \frac{2^{4(i_2 + i_3 + \cdots +
i_{\Lambda})}}{2^{(4\mu-2j)}}
\left\{\prod_{m=2}^{\Lambda}D^{i_m}_{\xi_m}
\left[\frac{1}{\xi_m^{2\mu-2I_{m+1}}} \left(
\frac{1}{AQ_2Q_1}+\cdots+ \frac{\xi_2\cdots
\xi_{\Lambda-1}}{AQ_{\Lambda}Q_{\Lambda-1}}
\right)^{2\mu-j} \right]\right\}_{(\xi_\rightarrow 1)}\nonumber \\
\noalign{\bigskip}
 &&\left(\frac{1}{AQ_0Q_1}\right)^j
\left\{\binom{I_2}{\lambda}a^{2\lambda}_{2\lambda-j}\ Q^{4\lambda}_0
Q_1^{4(I_2-\lambda)}\right\}
\end {eqnarray}

To accomplish this calculation we change the order of the summations
and we apply the identities:

$$(4I_2-2j)! = 2^{(4I_2-2j)}\ (2I_2-j)!\ \pochh{1/2}{2I_2-j}$$
$$a^{2\lambda}_{2\lambda-j} = \binom{2\lambda}{2\lambda-j}\ \frac{\pochh{1/2}{2\lambda}}{\pochh{1/2}{2\lambda-j}}$$
$$\frac{\pochh{1/2}{2I_2-j}}{\pochh{1/2}{2\lambda-j}} =
\partial_{\varepsilon}^{2I_2-2\lambda}\ \left(\frac{1}{\varepsilon^{2\lambda-j+1/2}}\right)\Bigg |_{\varepsilon \rightarrow 1}$$
$$\frac{(2I_2-j)!}{(2\mu -j)!} = \int\ d\vartheta^{2\mu-2I_2}\ \left(\vartheta^{2I_2-j}\right)\Bigg |_{\vartheta \rightarrow 1}$$
$$\frac{2^{4(i_2 + i_3 + \cdots + i_{\Lambda})}}{2^{(4\mu-4I_2)}}\ = 1$$

We find:

\begin {eqnarray}
&& \sum_{\lambda = 0}^{I_2}\binom{I_2}{\lambda}\
\pochh{1/2}{2\lambda}\ Q^{4\lambda}_0 Q_1^{4(I_2-\lambda)}
\label{ap6}
\\ \noalign{\bigskip}\nonumber
 &&\sum_{j=0}^{2\lambda}\ \binom{2\lambda}{j}
 \left\{\partial_{\varepsilon}^{2I_2-2\lambda}\ \left(\frac{1}{\varepsilon^{2\lambda-j+1/2}}\right)\Big|_{\varepsilon \rightarrow 1}
 \right\}
 \left\{\int\ d\vartheta^{2\mu-2I_2}\ \left(\vartheta^{2I_2-j}\right)\Big|_{\vartheta \rightarrow
 1}\right\}\\ \noalign{\bigskip}\nonumber
&&\left(\frac{1}{AQ_0Q_1}\right)^j
\left\{\prod_{m=2}^{\Lambda}D^{i_m}_{\xi_m}
\left[\frac{1}{\xi_m^{2\mu-2I_{m+1}}} \left(
\frac{1}{AQ_2Q_1}+\frac{\xi_2}{AQ_3Q_2}+\cdots+
\frac{\xi_2\cdots\xi_{\Lambda-1}}{AQ_{\Lambda}Q_{\Lambda-1}}
\right)^{2\mu-j} \right]\right\}\Bigg |_{(all \xi_m\rightarrow 1)}
\end {eqnarray}

The derivative over variable $\varepsilon$ and inverse derivative
over variable $\vartheta$ are independent on the summation index $j$
and we can transfer them out of the sum. The summation over index
$j$ gives:

\begin {equation}
\sum_{j=0}^{2\lambda}\ \binom{2\lambda}{j}\ \left(\frac{\varepsilon
U}{\vartheta}\right)^j\ = \left(1+\frac{\varepsilon
U}{\vartheta}\right)^{2\lambda}
\end {equation}

where

$$U = \left(
\frac{1}{AQ_2Q_1}+\frac{\xi_2}{AQ_3Q_2}+\cdot\cdot\cdot+
\frac{\xi_2\cdot\cdot\cdot\xi_{\Lambda-1}}{AQ_{\Lambda}Q_{\Lambda-1}}
\right)^{-1} \left(\frac{1}{AQ_0Q_1}\right).$$

For summation over index $j$ in (\ref{ap6}) we obtain:

\begin {eqnarray}
&&\left\{\prod_{m=2}^{\Lambda}D^{i_m}_{\xi_m}
\left[\frac{1}{\xi_m^{2\mu-2I_{m+1}}} \left(
\frac{1}{AQ_2Q_1}+\frac{\xi_2}{AQ_3Q_2}+\cdot\cdot\cdot+
\frac{\xi_2\cdot\cdot\cdot\xi_{\Lambda-1}}{AQ_{\Lambda}Q_{\Lambda-1}}
\right)^{2\mu} \right]\right\} \\
\noalign{\bigskip}\nonumber &&\left\{\int\ d\vartheta^{2\mu-2I_2}\
\left(\vartheta^{2I_2}\right)\right\}
\left\{\partial_{\varepsilon}^{2I_2-2\lambda}\
\left(\frac{1}{\varepsilon^{2\lambda
+1/2}}\right)\right\}\left(1+\frac{\varepsilon
U}{\vartheta}\right)^{2\lambda}\Bigg |_{all \xi_m\rightarrow 1,
\varepsilon \rightarrow 1, \vartheta \rightarrow 1}
\end {eqnarray}

 We stress that the
object $U$ is proportional to $\triangle$. In the above relation we
proceed by evaluation of the derivative over variable $\varepsilon$.
We have:

\begin {eqnarray}
& &\left\{\partial_{\varepsilon}^{2I_2-2\lambda}\
\left(\frac{1}{\varepsilon^{2\lambda
+1/2}}\right)\left(1+\frac{\varepsilon
U}{\vartheta}\right)^{2\lambda}\right\}\Bigg |_{\varepsilon
\rightarrow 1} = \nonumber \\
\noalign{\bigskip} & &\sum_{i=0}^{\min{(2\lambda, 2I_2-2\lambda)}}\
(-1)^i\binom{2I_2-2\lambda}{i}\ \pochh{2\lambda +
1/2}{2I_2-2\lambda-i}\
\frac{(2\lambda)!}{(2\lambda-i)!}\left(1+\frac{\varepsilon
U}{\vartheta}\right)^{2\lambda-i}\left(\frac{
U}{\vartheta}\right)^i\Bigg |_{\varepsilon \rightarrow 1}
\label{ap5}
\end {eqnarray}

With help of identity:

\begin {eqnarray}
\pochh{2\lambda + 1/2}{2I_2-2\lambda-i} =
\frac{\binom{2I_2-1/2}{2I_2-2\lambda}(2I_2-2\lambda)!}{\binom{2I_2-1/2}{i}(i)!}
\end {eqnarray}
we converted (\ref{ap5}) to the final form:

\begin {eqnarray}
&&\left\{\partial_{\varepsilon}^{2I_2-2\lambda}\
\left(\frac{1}{\varepsilon^{2\lambda
+1/2}}\right)\left(1+\frac{\varepsilon
U}{\vartheta}\right)^{2\lambda}\right\}\Bigg |_{\varepsilon
\rightarrow 1} =
\binom{2I_2-1/2}{2I_2-2\lambda}(2I_2-2\lambda)!
\left(1+\frac{U}{\vartheta}\right)^{2\lambda} \label{ap4}
\\ \noalign{\bigskip}\nonumber
&&\sum_{i=0}^{MIN}\ (-1)^i\ \binom{MIN}{i}\ \binom{MAX}{i}\
\binom{2I_2-1/2}{i}^{-1}\ \left(\frac{U}{U+\vartheta}\right)^i \Bigg
|_{\vartheta \rightarrow 1}
\end {eqnarray}

where $MIN=\min{(2\lambda,\ 2I_2-2\lambda)}$, and
$MAX=\max{(2\lambda,\ 2I_2-2\lambda)}$.

The sum over index $i$ can be expressed by help of Gegenbauer
orthogonal polynomials $C_n^{\alpha}$ following the relation (see
\cite{prud}, vol. I, page 634):

\begin {eqnarray}
\sum_{i=0}^n\ (-1)^i\ \binom{n}{i}\ \binom{a-n+1/2}{i}\
\binom{a}{i}^{-1}\ x^k =
\binom{a}{2n}^{-1}\left(-\frac{x}{4}\right)^n\
C_{2n}^{1+a-2n}(\sqrt{1-1/x})
\end {eqnarray}

We insert (\ref{ap4}) into the relation (\ref{ap6}), and using the
identity:

$$\pochh{1/2}{2\lambda}\ (2I_2-2\lambda)!\
\binom{2I_2-1/2}{2I_2-2\lambda} = \pochh{1/2}{2I_2}$$ we find:

\begin {eqnarray}
& &\pochh{1/2}{2I_2}\ \left\{\prod_{m=2}^{\Lambda}D^{i_m}_{\xi_m}
\left[\frac{1}{\xi_m^{2\mu-2I_{m+1}}} \left(
\frac{1}{AQ_2Q_1}+\frac{\xi_2}{AQ_3Q_2}+\cdot\cdot\cdot+
\frac{\xi_2\cdot\cdot\cdot\xi_{\Lambda-1}}{AQ_{\Lambda}Q_{\Lambda-1}}
\right)^{2\mu} \right]\right\}\\ \noalign{\bigskip} &
&\sum_{\lambda=0}^{I_2}\binom{I_2}{\lambda}\
 \binom{2I_2-1/2}{2 MIN}^{-1}\ Q_0^{4\lambda}Q_1^{4(I_2-\lambda)}\nonumber \\
\noalign{\bigskip} & &\left\{\int\ d\vartheta^{2\mu-2I_2}\
\left(\vartheta^{2I_2-2\lambda+MIN}\right)\right\}\
(U+\vartheta)^{2\lambda-MIN}\
\left(-\frac{U}{4\vartheta}\right)^{MIN}\ C_{2
MIN}^{MAX-MIN+1/2}(\sqrt{-\vartheta/U})\Bigg |_{\vartheta
\rightarrow 1}\nonumber
\end {eqnarray}

We see that the term $\pochh{1/2}{2I_2}$ determine the force of
divergence. This means, that the leading term have this term as
great as possible and this situation set if $I_2 = \mu$. In this
case all indices $i_m = 0$. This means that the derivative operators
$D^{i_m}_{\xi_m}$ have no effects in this case. The argument of the
Gegenbauer polynomials are going to infinity as
$\sqrt{1/\triangle}$. For $\triangle$ finite, but $\triangle
\rightarrow 0$  we can use the relation (see Szeg\H{o} \cite{sego})

$$ \lim_{x \rightarrow \infty}\ x^{-n}\ C_n^{\lambda}(x) = 2^n\
\binom{n+\lambda-1}{n}.$$ Inserting this into above relation we have
approximatively:

\begin {eqnarray}
&&\pochh{1/2}{2I_2}\ \left(
\frac{1}{AQ_2Q_1}+\frac{1}{AQ_3Q_2}+\cdot\cdot\cdot+
\frac{1}{AQ_{\Lambda}Q_{\Lambda-1}} \right)^{2\mu} \
\sum_{\lambda=0}^{I_2}\binom{I_2}{\lambda} \
Q_0^{4\lambda}Q_1^{4(I_2-\lambda)} \\
\noalign{\bigskip}\nonumber &&\left\{ \int\ d\vartheta^{2\mu-2I_2}\
\left(\vartheta^{2I_2-2\lambda+MIN}\right)\right\}\
(U+\vartheta)^{2\lambda-MIN}
\end {eqnarray}

 Taking into account,
that $I_2 = \mu$, the inverse derivative over variable $\vartheta$
have no effect and we red:

$$\left\{ \int\ d\vartheta^{2\mu-2I_2}\
\left(\vartheta^{2I_2-2\lambda+MIN}\right)\right\}\
(U+\vartheta)^{2\lambda-MIN} \rightarrow
\vartheta^{2\mu-2\lambda+MIN}\ (U+\vartheta)^{2\lambda-MIN}\Bigg
|_{\vartheta\rightarrow 1}$$

Taking into account that $U \sim \triangle$ we have:

$$\left\{\ms C(\Lambda)^{2\mu}\right\}_{2\mu,2\mu}\; =\; \pochh{1/2}{2\mu}\left( \frac{1}{AQ_2Q_1}+\frac{1}{AQ_3Q_2}+\cdot\cdot\cdot+
\frac{1}{AQ_{\Lambda}Q_{\Lambda-1}} \right)^{2\mu}
(Q_0^4+Q_1^4)^{\mu}\ $$

For finite but $\triangle \rightarrow 0$ we approximatively find:
$$\left\{\ms C(\Lambda)^{2\mu}\right\}_{2\mu,2\mu}\; \sim \ 2^{\mu}\
\pochh{1/2}{2\mu}\left(\frac{\tanh{(\gamma
\tau)}}{\triangle\gamma}\right)^{2\mu}$$

Finally, for the leading divergent terms for $\mu \rightarrow
\infty$ of the asymptotic series for $\mathcal{S}_{\Lambda}$  in
quasi-continuum relation we have:

\begin {equation}
\frac{(-1)^{\mu}a^{\mu}}{\mu!\;c^{2\mu}}\; \triangle^{\mu}2^{\mu}\
\pochh{1/2}{2\mu}\left(\frac{\tanh{(\gamma
\tau)}}{\gamma}\right)^{2\mu}
\end {equation}

\end {document}